\documentclass[10pt,journal]{IEEEtran}
\usepackage{amsmath,amssymb}
\usepackage{graphicx}
\usepackage{verbatim}
\usepackage{multirow,cite}
\usepackage{epsfig}
\usepackage{algorithm,algorithmic}

\def\d{\downarrow}
\def\u{\uparrow}

\begin{document}
\title{Improved Analytical Delay Models for RC-Coupled Interconnects}

\author{Feng Shi, Xuebin~Wu, and Zhiyuan~Yan

  \thanks{Feng Shi, Xuebin Wu, and Zhiyuan
    Yan are with the Department of ECE,
    Lehigh University, PA 18015, USA. E-mails:\{fes209, xuw207,
     yan\}@lehigh.edu.}}


%
%
%
%
%

\maketitle
\begin{abstract}
As the process technologies scale into deep submicron region, crosstalk delay is becoming increasingly severe, especially for global on-chip buses. To cope with this problem, accurate delay models of coupled interconnects are needed.
In particular, delay models based on analytical approaches are desirable, because they not only are largely transparent to technology, but also explicitly establish the connections between delays of coupled interconnects and transition patterns, thereby enabling crosstalk alleviating techniques such as crosstalk avoidance codes (CACs).
Unfortunately, existing analytical delay models, such as the widely cited model in~\cite{Sot01}, have limited accuracy and do not account for loading capacitance. In this paper, we propose analytical delay models for coupled interconnects that address these disadvantages.
By accounting for more wires and eschewing the Elmore delay, our delay models achieve better accuracy than the model in \cite{Sot01}.
\end{abstract}

\section{Introduction}
\label{sec:intro}
Crosstalk caused by coupling capacitance between adjacent wires leads to additional delay to multi-wire buses. As the process technologies scale into deep submicron region,
coupling capacitance between adjacent wires and hence crosstalk delays increase greatly. According to the International Technology Roadmap
of Semiconductors (ITRS) \cite{ITRS}, gate delay \textbf{decreases} with scaling, while
global wire delay \textbf{increases}. Hence, the crosstalk delay problem is becoming increasingly severe in VLSI designs, especially for global on-chip buses, and will become the performance bottleneck in many high-performance VLSI designs.

This paper focuses on analytical delay models applicable to general coupled interconnects. Although various delay models of interconnects have been proposed in the literature (see, for example, \cite{Kay98,Pil98,Alp01,Sot01,Sot02,Liu02,Dav002,Abo02,Tu06,Mol09}), few are comparable to our work in this paper. Some delay models (see, for example, \cite{Kay98,Alp01,Liu02,Mol09,Pil98,Abo02}) do not consider crosstalk from adjacent wires. Furthermore, most previously proposed delay models are based on numerical approaches (see, e.g., \cite{Kay98,Alp01,Liu02,Dav002,Tu06,Mol09,Pil98,Abo02}). They can achieve high accuracy, but they have several drawbacks. First, they sometimes lead to lookup tables of delays from any initial state to any next state (see, for example, \cite{Mol09}), which are often bulky and difficult to obtain and use. Second, numerical approaches in \cite{Kay98,Pil98,Alp01,Liu02,Abo02} are technology-dependent and their delays often depend on many parameters. Hence these approaches have poor portability and are not applicable to general cases. Third, the delays obtained by the numerical approach offer little insight, and are not conducive to technology-independent crosstalk alleviation techniques such as crosstalk avoidance codes (CACs) (see, for example, \cite{Dua01,Vic01,Sri07,Wu09}). Fourth, numerical approaches often have very high complexities. In contrast, analytical approaches are advantageous in these aspects. Analytical approaches depend on few technology parameters, and hence they are largely technology independent. Furthermore, analytical approaches illustrate the connection between delays of coupled interconnects and transition patterns, thus enabling us to design CACs. Finally, analytical approaches have very low computational complexities.
A widely cited analytical delay model proposed by Sotiriadis \emph{et al.} \cite{Sot01,Sot02}, which uses the similar methodology to that in \cite{SC_TVLSI02}, appears to be the most comparable previous delay model to our work in this paper.


Based on the model in \cite{Sot01,Sot02}, the delay of the $k$-th wire ($k \in \left\{1,2,\cdots,m \right\}$) of an $m$-bit bus is given by
\begin{equation}
  T_k = \left\{
    \begin{array}{ll}
      \tau_0[(1+\lambda)\Delta_1^2-\lambda\Delta_1\Delta_2], & k=1\\
      \tau_0[(1+2\lambda)\Delta_k^2-\lambda\Delta_k(\Delta_{k-1}+\Delta_{k+1})],
      & k\neq 1,m\\
      \tau_0[(1+\lambda)\Delta_m^2-\lambda\Delta_m\Delta_{m-1}],& k=m,\\
    \end{array}
  \right.
  \label{Eq:1}
\end{equation}
where $\lambda$ is the ratio of the coupling capacitance between
adjacent wires and the ground capacitance of each wire,
$\tau_0$ is the intrinsic delay of a transition on a single wire,
and $\Delta_k$ is $1$ for $0\rightarrow 1$ transition, $\textnormal{-}1$ for
$1\rightarrow 0$ transition, or $0$ for no transition on the $k$-th
wire. We observe that in this model, the delay of the $k$-th wire
depends on the transition patterns of wires $k-1$, $k$, and $k+1$ only. Since all possible values of $T_k$ in Eq.~(\ref{Eq:1}) are $(1+i\lambda)\tau_0$ for $i \in \{0,1,2,3,4\}$, all transition patterns on wires $k-1$, $k$, and $k+1$ can be divided into five classes according to their corresponding $i$. These five classes are denoted as $iC$ for $i \in \{0,1,2,3,4\}$ (this classification was also used in \cite{Dua01}). Based on this model, various CACs (see, for example,  \cite{Dua01,Vic01,Sri07,Wu09}) have been proposed, based on the central idea of achieving a reduced delay by limiting transition patterns over the bus, at the expense of additional wires.

However, the model in \cite{Sot01} have two significant drawbacks. First, the model in \cite{Sot01} has limited accuracy. In a bus with more than three wires, the simulated wire delay for $0C$ transition patterns is much larger than $\tau_0$, the delay of $0C$ given by (\ref{Eq:1}). This implies that the scheme that uses two shield wires with the same transition to achieve a delay of $\tau_0$ (see, for example, \cite{Dua04}) will be ineffective. Our simulation results also show that the delays of other classes of transition patterns given by Eq.~(\ref{Eq:1}) have limited accuracy as well. This is partially because of the model's dependence on only three wires.
Also, the model in \cite{Sot01} uses in its derivation the Elmore delay, which tends to overestimate the delay \cite{Elm48,Gup97}.

The second drawback of the model in \cite{Sot01} is that it does not account for the loading capacitance. It has been shown that the loading capacitance is crucial in real practice and can affect the total delay for all patterns.

Addressing these disadvantages for the model in \cite{Sot01}, in this paper we propose analytical delay models for coupled interconnects. Our delay models first derive closed-form expressions of the signals on the bus via a distributed RC model, and then approximate the wire delays by evaluating these closed-form expressions.
Our delay models differ from the model in \cite{Sot01} in three aspects. First, in our delay models, we eschew the Elmore delay used in the model in \cite{Sot01}. Then, we consider either three wires or five wires in our delay models for improved accuracy. Due to these two differences, our models have significantly improved accuracy than the model in \cite{Sot01}. Finally, we take into account the buffer effects (driver resistance and loading capacitance). Our delay models also maintain the simplicity of the model in \cite{Sot01}, and the transition patterns are divided into several categories based on their delays. Hence, our delay models are easy to use and conducive to the design of CACs.
Although our delay models consider adjacent three and five wires in this paper, our models are applicable to buses of any number of wires.

Simulation results show that our delay models offer significant advantages than the model in \cite{Sot01}. Our simulations results fall into three scenarios. First, we compare the delays produced by our model and the model in \cite{Sot01} with the simulated delays for three- or five-wire buses. This is motivated by partial coding schemes (see, e.g., \cite{Dua01}, \cite{Vic01}, and \cite{Sri07}), which divide a wide bus into sub-buses with a few wires and separate them by shielding wires.
Second, we obtain extensive simulation results for 17- and 33-wire buses assuming arbitrary transition patterns. Third, we assume the transition patterns are limited to those of CACs. In all three scenarios, our five-wire delay model is much more accurate than the model in \cite{Sot01}.

With the scaling of technologies, the inductance is becoming significant and impacts the signals on the bus greatly. Due to the coupling effect of inductance, the worst-case patterns for an RLC modeled bus are quite different from that of an RC modeled bus~\cite{Tu06}. Hence, the CAC design methodology would change greatly due to the inductance effect.
However, our delay models do not consider the inductance effect  for two reasons.
First, it seems difficult to derive a closed-form expression of the signals on the bus based on the RLC model. Hence, our methodology cannot be easily adapted from the distributed RC model to an RLC model. Second, according to the criteria in \cite{Ism99}, the inductance effect is significant in some cases, but are negligible in other cases. Specifically, the range of significance of the inductance effect is given by $\frac{t_r}{2\sqrt{lc}}<x<\frac{2}{r}\sqrt{\frac{l}{c}}$ \cite{Ism99}, where $x$ is the length of the wire, $t_r$ the input transition time, and $r$, $l$, and $c$ the resistance, inductance, and capacitance per unit length, respectively.
According to \cite{CSLJ_JLPE08}, the inductance effect is not negligible for very deep submicron technologies and extremely long wires.
In current industry applications, the on-chip inductance effect is still insignificant. This conclusion was also confirmed by other works: the 16-bit, 32Gb/s, 5mm-long bus and 8-bit, 16Gb/s, 10mm-long bus in \cite{Zha09} show that the distributed RC model is sufficiently accurate for these high-speed long interconnects. In our work, our delay models are derived based on 5mm-long buses under a 45nm technology, where inductance effect is negligible.

The rest of the paper is organized as follows. In Section~\ref{sec:model}, we propose our delay models. The delay models are also modified to account for the buffer effects.
In Section~\ref{sec:simulation}, we present extensive simulation results for our delay models. Concluding remarks are provided in Section~\ref{sec:conclusion}.

\section{Delay model}
\label{sec:model}

In this section, we first present the system model, where switching instants of all wires in the bus are assumed simultaneous. For three-wire and five-wire buses, we then derive closed-form expressions for outputs of the bus, and finally approximate their delays and compare them with those by the model in \cite{Sot01}.

\subsection{System model}

In this paper, we focus on global interconnects connecting different modules for communication, such as data and address buses, and use the distributed RC model for interconnect modeling. For simplicity, we assume regular interconnects, which have uniformly distributed parameters and are paralleled routed in the same metal layer without turnings. Hence, the interconnects are modeled as transmission lines, which can be characterized by the telegrapher's equations.
For complex interconnect structures with jumps and turnings, additional resistance due to vias and unequal length of wires should be included, which makes the interconnect behavior more complicated. However, crosstalk delays are expected to increase due to the additional resistance.
The partially coupled buses are more complex and hence are not considered in this paper. We plan to investigate this in our future work.
The capacitance between non-adjacent wires is negligible compared with capacitance between adjacent wires, since the capacitive coupling effect is a short range effect~\cite{SC_TVLSI02}.
The distributed RC model is often used to approximate the buses~\cite{Rab96}. Although the closed-form expressions of the signals on the bus via a distributed RC model are sums of infinite terms, usually sums of the two most significant terms provide a very close approximation of signals on the bus \cite{Sak93}.

\begin{figure}[!tb]
\begin{minipage}[b]{1.0\linewidth}
  \centering
 \centerline{\epsfig{figure=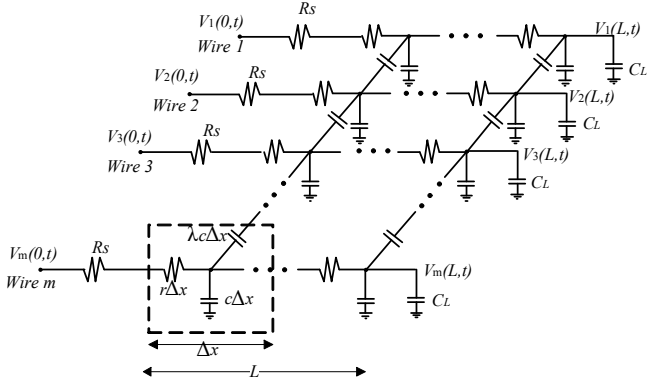,width=8.5cm}}
\end{minipage}
\caption{A distributed RC model of an $m$-wire bus.}
\label{fig:3bus}
\end{figure}

The distributed RC model of an $m$-wire bus is shown in Fig.~\ref{fig:3bus}, where $V_i(x,t)$ denotes the transient signal at a position $x$ along wire $i$ for $i \in \{1,2,\cdots,m\}$, $r$ and $c$ denote the resistance and capacitance per unit length, respectively. Also, $\lambda_c$ denotes the coupling capacitance per unit length between two adjacent wires. The output resistance of a driver is approximated as a linear resistor, $R_S$, and the loading due to a receiver is modeled as a capacitance, $C_L$. In this work, we focus on a uniformly distributed bus and hence assume the parameters $r$, $c$, and $\lambda$ are the same for all the wires.

We use the 50\% delay, which is defined as the time difference between the respective instants when the input signal and corresponding output signal cross 50\% of the supply voltage $V_{dd}$. According to \cite{Kah99}, the delays of global interconnects are slightly affected by the slew rate. Since this work focuses on global interconnects, we ignore input slew and assume ideal step signals are applied on the bus directly. In this paper, we use the same classification $iC$ for $i=0,1,2,3,4$ in \cite{Sot01} and focus on the worst 50\% delay of any wire for all classes to formulate our delay models.
We consider the closest neighbors for crosstalk, since farther wires have weaker coupling effects.
In Section~\ref{sec:3m}, we first focus on internal wire, wire 2 in a three-wire model, to account for most adjacent two wires (one wire to the left and one to the right). In Section~\ref{sec:5w}, we focus on internal wire, wire 3 in a five-wire model, to account for most adjacent four wires (two wires to the left and two to the right).
In Section~\ref{sec:boundary}, we derive the delay for boundary wires, wires 1, 2, 4, and 5 in a five-wire model.
In Section~\ref{sec:application}, we show how to identify the worst-case delays among all wires for a wide bus via a shift window scheme.


In this section, we first derive delay models by assuming that the buffer effects (driver resistance and loading capacitance) are negligible. This is an important case since the propagation delay is characterized only by the distributed interconnects. Then, in Section~\ref{sec:revised}, we modify the delay models to account for the buffer effects, which are crucial in real practice. It has been shown that the buffer effects would increase the total delay for all patterns.

Below we first investigate the case $m=3$ and then extend our results to the case $m=5$. There are two reasons for studying the three-wire model. First and foremost, the derivation of our five-wire model is based on the three-wire model. Second, our three-wire model is more accurate than our five-wire model for buses with only three wires, which is of interest for partial coding schemes (see, e.g., \cite{Dua01}, \cite{Vic01}, and \cite{Sri07}).
We use $T^{iC}_m$ to denote the worst delay of the middle wire (wire $\frac{m+1}{2}$) of an $m$-wire bus for all $iC$ patterns.

\subsection{Internal wires for three-wire model}
\label{sec:3m}

In~\cite{Sak93}, the crosstalk of two coupled lines was described by partial differential equations (PDEs), and a technique for decoupling these highly coupled PDEs was introduced by using eigenvalues and corresponding eigenvectors.
Using the same technique as in \cite{Sak93}, we obtain the differential equations describing a three-wire bus with length $L$:
\begin{equation}
\frac{\partial^2}{\partial x^2}\mathbf{V}(x,t)
=\mathbf{RC}\frac{\partial}{\partial t}\mathbf{V}(x,t),
\label{Eq:2}
\end{equation}
where $\mathbf{V}(x,t)=[V_1(x,t)\;V_2(x,t)\;V_3(x,t)]^T$ and $V_i(x,t)$ denotes the voltage of wire $i$ at distance $x$ ($0\le x \le L$) at time $t$ for $i=1,2,3$,
$\mathbf{R}=\mbox{diag}\{r \; r \; r\}$, and $\mathbf{C}=c\left[ \begin{array}{ccc} 1+\lambda & -\lambda & 0 \\ -\lambda & 1+2\lambda & -\lambda\\
0 & -\lambda & 1+\lambda \end{array}\right]$.

The boundary conditions are given by
\begin{equation*}
\left\{ \begin{array}{l}
V_i(0,t)=V_i^p-(V_i^p-V_i^f) u(t) \\
I_i(L,t)=0
\end{array} \mbox{ for } i=1,2,3 \right.
\end{equation*}
where $V^p_i$ and $V^f_i$ denote the initial and final voltages of the transition on wire $i$, respectively.


We find the three eigenvalues of $\mathbf{C}/c$, $p_1=1$, $p_2=(1+\lambda)$, and $p_3=(1+3\lambda)$, and their corresponding eigenvectors $\mathbf{e}_i$'s, $[1 \, 1 \, 1]^T$, $[1 \, 0 \, \textnormal{-}1]^T$, and $[\textnormal{-}1 \, 2 \, \textnormal{-}1]^T$, respectively. Hence, Eq.~(\ref{Eq:2}) is transformed to
\begin{equation}
\frac{\partial^2}{\partial x^2}U_i(x,t)=rcp_i\frac{\partial}{\partial t} U_i(x,t) \mbox{ for }i=1,2,3,\label{Eq2}
\end{equation}
where $U_i(x,t)=\mathbf{V}^T(x,t) \mathbf{e}_i$ for $i=1,2,3$. So $U_1(x,t)=V_1(x,t)+ V_2(x,t) + V_3(x,t)$, $U_2(x,t)=V_1(x,t)-V_3(x,t)$, and $U_3(x,t)=2V_2(x,t)-V_1(x,t)-V_3(x,t)$.

Applying Laplace transform on both sides of Eq.~(\ref{Eq2}), we have
\begin{equation}
\frac{\partial^2}{\partial x^2}U_i(x,s)=rcp_i[s U_i(x,s) - U_i(x,0)] \mbox{ for }i=1,2,3. \label{Eq1}
\end{equation}

Using appropriate initial conditions, we solve Eq.~(\ref{Eq1}) for $U_i(x,t)$ and obtain $V_2(L,t)=\frac{1}{3}[U_1(L,t)+U_3(L,t)]$.
By solving $V_2(L,t)=0.5V_{dd}$, we can approximate the 50\% delay of a three-wire bus for different transition patterns.

In this paper, we use ``$\uparrow$" to denote a transition from 0 to the supply voltage $V_{dd}$ (normalized to 1), ``-" no transition, and ``$\downarrow$" a transition from $V_{dd}$ to 0.

For $0C$ pattern $\u\u\u$, the output of wire 2 is given by~\cite{Sak93} $V_2(L,t)=1 + \sum\limits^\infty_{n=1}\frac{(\textnormal{-}1)^n}{\frac{\pi}{4}(2n-1)} e^{-\frac{t}{\tau}(2n-1)^2}$, where $\tau_0=\frac{rcL^2}{2}$, and $\tau=\frac{8}{\pi^2}\tau_0$.

For the 50\% delay, keeping only the first exponential term is accurate enough. So we have $V_2(L,t)\doteq 1- \frac{4}{\pi}e^{-\frac{t}{\tau}}$.
Similarly, we keep only the first exponential term as the solution for other cases.
Solving $V_2(L,t)=0.5$, we have $T^{0C}_3\doteq\left(\ln{\frac{8}{\pi}}\right)\tau$.
Similarly, the closed-form expressions of wire 2 and approximate delays for other classes are derived and summarized in Table~\ref{tab:3wire}, where $T^{iC}_3$ the approximate delay for $iC$ pattern by our three-wire model.

\begin{table}[!thb]
\caption{Closed-form expressions of signal on wire 2 and approximate delays in a three-wire bus ($V_2(L,t) = 1-A_1 e^{-\frac{t}{\tau}} - A_2 e^{-\frac{t}{(1+3\lambda)\tau}}$, $\tau_0=\frac{rcL^2}{2}$, and $\tau=\frac{8}{\pi^2}\tau_0$.)}\label{tab:3wire}
\begin{center}
\begin{tabular}{|c|c|c|c|c|}
\hline
\multirow{2}{*}{$iC$} & Worst & \multicolumn{2}{|c|}{Coeffs. of $V_2(L,t)$} & \multirow{2}{*}{$T_3^{iC}$}\\
\cline{3-4}
& Pattern & $A_1$ & $A_2$ & \\
\hline
$0C$ & $\u\u\u$ & $\frac{4}{\pi}$ & 0 & $\left(\ln{\frac{8}{\pi}}\right)\tau$ \\
\hline
$1C$ & $\u\u$- & $\frac{8}{3\pi}$ & $\frac{4}{3\pi}$ & $\left(\ln{\frac{16}{\pi}}\right)\tau$\\
\hline
$2C$ & -$\u$- & $\frac{4}{3\pi}$ & $\frac{8}{3\pi}$ & $\left(\ln{\frac{16}{3\pi}}\right)(1+3\lambda)\tau$ \\
\hline
$3C$ & $\d\u$- & 0 & $\frac{4}{\pi}$ & $\left(\ln{\frac{8}{\pi}}\right)(1+3\lambda)\tau$ \\
\hline
$4C$ & $\d\u\d$ & $\textnormal{-}\frac{4}{3\pi}$ & $\frac{16}{3\pi}$ & $\left(\ln{\frac{32}{3\pi}}\right)(1+3\lambda)\tau$ \\
\hline
\end{tabular}
\end{center}
\end{table}

\subsection{Internal wires for five-wire model}\label{sec:5w}
To further improve the accuracy of delay, we include two extra adjacent wires to approximate the delay by considering the influences of all five wires. Each wire has three kinds of transition: $\uparrow$, -, and $\downarrow$. Hence, for such a five-wire bus, there are $3^5$ transition patterns. To maintain the simplicity of our models, we still divide them into five classes ($iC, \; i\in \{0,1,2,3,4\}$) based on the transition patterns of middle three wires (wires 2, 3, and 4). Hence, there are nine different transition patterns for each pattern of the same class.

Since the interconnect is a linear system, any pattern can be decomposed into a combination of patterns with transitions on a single wire. For example, $\u\u\u\d$- is decomposed as ($\u$- - - -) +  (-$\u$- - -) + (- -$\u$- -) + (- - -$\d$-). The delay expression of the middle wire impacted by any pattern is given by a summation of effects of individual wires on the middle wire.
However, this approach would result in expressions that are hard to analyze. Instead, we propose to group these individual wires to form some special patterns, which can be analyzed easily.\\

\textbf{Definition 1}: Reducible transition pattern (RTP)

An RTP in the five-wire model is defined as a transition pattern that can be reduced to a transition pattern in the three-wire model. The set \{$\uparrow \uparrow \uparrow \uparrow \uparrow$, $\downarrow\downarrow\downarrow\downarrow\downarrow$, $\downarrow$-$\uparrow$-$\downarrow$, $\uparrow$-$\downarrow$-$\uparrow$\} is the set of RTPs for the five-wire model.

For the transition $\uparrow \uparrow \uparrow \uparrow \uparrow$ (similarly for $\downarrow \downarrow \downarrow \downarrow \downarrow$), all wires have the same transitions. There are no coupling capacitance between any two adjacent wires. So the expression of wire 3 is approximated by
$V_3(L,t) \doteq 1-\frac{4}{\pi}e^{-\frac{t}{\tau}}$ and the delay is approximated by $\left(\ln{\frac{8}{\pi}}\right)\tau$. For the transition $\downarrow$-$\uparrow$-$\downarrow$ (similarly for $\uparrow$-$\downarrow$-$\uparrow$), wires 2 and 4 can be approximated as ground wires in the five-wire bus, since wire 1 (or 5) and wire 3 have opposite transitions. For wire 3, the five-wire pattern is equivalent to a three-wire pattern $\downarrow \uparrow \downarrow$, where the equivalent coupling capacitor between wire 1 (or 5) and wire 3 is equal to two capacitors in series between wires 1 and 2, and wires 2 and 3 (or wires 3 and 4, and wires 4 and 5).
Hence, the equivalent coupling factor between wire 1 (or 5) and wire 3 is approximated as $\frac{\lambda}{2}$ per unit length (that is, the ratio of the coupling capacitance and the loading capacitance is $\frac{\lambda}{2}$). The expression of wire 3 is approximated by $V_3(L,t) \doteq 1+\frac{4}{3\pi} e^{-\frac{t}{\tau}} - \frac{16}{3\pi} e^{-\frac{t}{(1+\frac{3}{2}\lambda)\tau}}$, and the delay is approximated by $\ln (\frac{16}{3\pi})(1+\frac{3}{2}\lambda)\tau$.\\

\textbf{Definition 2}: Single transition pattern (STP)

An STP is defined to be a transition pattern with transitions on only one wire. For our five-wire model, we focus on the set of STPs with transitions on wire 2 or 4, \{-$\uparrow$-~-~-, -$\downarrow$-~-~-, -~-~-$\uparrow$-, -~-~-$\downarrow$-\}.

The expressions of wire 3 can be approximated by considering wires 2, 3, and 4 as a three-wire model. Let $V^{i}_j(x,t)$ denote the signal on wire $j$ due to coupling from wire $i$. For example, by ignoring coupling from wires 1 and 5 in -$\uparrow$- - -, the output of wire 3 is approximated by  $V^{2}_{3}(L,t)\doteq \textnormal{-}\frac{4}{3\pi}e^{-\frac{t}{\tau}}+\frac{4}{3\pi}e^{-\frac{t}{(1+3\lambda)\tau}}$, which is obtained by considering only wires 2, 3, and 4.\\

We propose the following approaches to derive the delay of the five-wire bus.
\begin{itemize}
\item We first \textbf{decompose} the worst pattern in each class into a combination of an RTP and STP(s).
\item Then we combine the expressions of the RTP and STP(s) for the middle wire based on the conclusion of our three-wire model.
\item Finally, we evaluate the expression of the middle wire to approximate its delay.
\end{itemize}

Since the performance is limited by the worst-case delay in each class, we need to approximate the delays of only the worst patterns in each class.
We use simulation to identify the worst patterns in all classes.
The worst patterns for $0C$ to $4C$ are given by $\downarrow \uparrow \uparrow \uparrow \downarrow$, $\downarrow \uparrow \uparrow$-$\downarrow$, $\downarrow$-$\uparrow$-$\downarrow$, $\uparrow \downarrow \uparrow$-$\uparrow$, and $\uparrow \downarrow \uparrow \downarrow \uparrow$, respectively (assuming the middle wire has an upward transition). With RTPs and STPs, we decompose the worst pattern in each class as shown in Table~\ref{tab:decomp}.
\begin{table}
\caption{Decomposition of worst-case patterns in the five-wire model.} \label{tab:decomp}
\begin{center}
\begin{tabular}{|c|c|l|}
\hline
$iC$ & Worst pattern & Decomposition\\
\hline
$0C$ & $\downarrow \uparrow \uparrow \uparrow \downarrow$ & ($\downarrow$-$\uparrow$-$\downarrow$)+(-$\uparrow$- - -)+( - - -$\uparrow$-)\\
\hline
$1C$ & $\downarrow$-$\uparrow \uparrow \downarrow$ & ($\downarrow$-$\uparrow$-$\downarrow$)+(- - -$\uparrow$-)\\
\hline
$2C$ & $\downarrow$-$\uparrow$-$\downarrow$ & ($\downarrow$-$\uparrow$-$\downarrow$)\\
\hline
$3C$ & $\uparrow$-$\uparrow \downarrow \uparrow$ & ($\uparrow \uparrow \uparrow \uparrow \uparrow$)+(- - -$\downarrow$-)+ (- - -$\downarrow$-) + (-$\downarrow$- - -)\\
\hline
$4C$ & $\uparrow \downarrow \uparrow \downarrow \uparrow$ & ($\uparrow \uparrow \uparrow \uparrow \uparrow$)+(-$\downarrow$- - -)+(-$\downarrow$- - -)+(- - -$\downarrow$-)+(- - -$\downarrow$-)\\
\hline
\end{tabular}
\end{center}
\end{table}

The closed-form expressions of wire 3 and approximate delays for all classes in a five-wire bus are derived and summarized in Table~\ref{tab:5wire}, where $T^{iC}_5$ the approximate delay for $iC$ pattern by our three-wire model.

\begin{table}[!t]
\caption{Closed-form expressions of signal on wire 3 and approximate delays in a five-wire bus ($V_3(L,t) = 1-A_3 e^{-\frac{t}{\tau}} - A_4 e^{-\frac{t}{(1+\frac{3}{2}\lambda)\tau}} - A_5 e^{-\frac{t}{(1+3\lambda)\tau}}$, $\tau_0=\frac{rcL^2}{2}$, and $\tau=\frac{8}{\pi^2}\tau_0$.)}\label{tab:5wire}
\begin{center}
\begin{tabular}{|c|c|c|c|c|c|}
\hline
\multirow{2}{*}{$iC$} & Worst & \multicolumn{3}{|c|}{Coeffs. of $V_3(L,t)$} & \multirow{2}{*}{$T_5^{iC}$}\\
\cline{3-5}
& Pattern & $A_3$ & $A_4$ & $A_5$ & \\
\hline
$0C$ & $\d\u\u\u\d$ & $\frac{4}{3\pi}$ & $\frac{16}{3\pi}$ & $\textnormal{-}\frac{8}{3\pi}$ & $0.165(1+3\lambda)\tau$ \\
\hline
$1C$ & $\d\u\u$-$\d$ & 0 & $\frac{16}{3\pi}$ & $\textnormal{-}\frac{4}{3\pi}$ & $0.384(1+3\lambda)\tau$\\
\hline
$2C$ & $\d$-$\u$-$\d$ & $\textnormal{-}\frac{4}{3\pi}$ & $\frac{16}{3\pi}$ & 0 & $\left(\ln{\frac{32}{3\pi}}\right)(1+\frac{3}{2}\lambda)\tau$ \\
\hline
$3C$ & $\u\d\u$-$\u$ & 0 & 0 & $\frac{4}{\pi}$ & $\left(\ln{\frac{8}{\pi}}\right)(1+3\lambda)\tau$ \\
\hline
$4C$ & $\u\d\u\d\u$ & $\textnormal{-}\frac{4}{3\pi}$ & 0 & $\frac{16}{3\pi}$ & $\left(\ln{\frac{32}{3\pi}}\right)(1+3\lambda)\tau$ \\
\hline
\end{tabular}
\end{center}
\end{table}

\subsection{Boundary wires}
\label{sec:boundary}
In the previous derivation, we focus on middle wires and consider four neighboring wires (two to the left and two to the right) for crosstalk. In this section, we derive delay models to account for the boundary wires of an $m$-wire bus (wires 1, 2, $m-1$, and $m$). For wire 1 (wire $m$), we consider wires 2 and 3 to the right (wires $m-2$ and $m-1$ to the left) for crosstalk, and use the same classification as in Eq.~(\ref{Eq:1})~\cite{Sot01}. Note that for wires 1 and $m$, there are only three classes of patterns, $0C$, $1C$, and $2C$.
With the similar technique, the closed-form expressions of wire 1 (wire $m$) and approximate delays for all classes are derived and summarized in Table~\ref{tab:wire1}, where $T_{b1}^{iC}$ is the approximate delay for $iC$ pattern.
For wire 2 (wire $m-1$), we consider wire 1 to the left and wires 3 and 4 to the right (wires $m-3$, $m-2$ to the left and wire $m$ to the right) for crosstalk. Similarly, the closed-form expressions of wire 2 (wire $m-1$) and approximate delays for all classes are derived and summarized in Table~\ref{tab:wire2}, where $T_{b2}^{iC}$ is the approximate delay for $iC$ pattern.

\begin{table}[!t]
\caption{Closed-form expressions of signal on wire 1 and approximate delays ($V_1(L,t) = 1-A_6 e^{-\frac{t}{\tau}} - A_7 e^{-\frac{t}{(1+\lambda)\tau}} - A_8 e^{-\frac{t}{(1+3\lambda)\tau}}$, $\tau_0=\frac{rcL^2}{2}$, and $\tau=\frac{8}{\pi^2}\tau_0$.)}\label{tab:wire1}
\begin{center}
\begin{tabular}{|c|c|c|c|c|c|}
\hline
\multirow{2}{*}{$iC$} & Worst & \multicolumn{3}{|c|}{Coeffs. of $V_1(L,t)$} & \multirow{2}{*}{$T_{b1}^{iC}$}\\
\cline{3-5}
& Pattern & $A_6$ & $A_7$ & $A_8$ & \\
\hline
$0C$ & $\u\u\d$ & $\frac{4}{3\pi}$ & $\frac{4}{\pi}$ & $\textnormal{-}\frac{4}{3\pi}$ & $0.783(1+\lambda)\tau$ \\
\hline
$1C$ & $\u$-$\d$ & 0 & $\frac{4}{\pi}$ & 0 & $(\ln \frac{8}{\pi})(1+\lambda)\tau$\\
\hline
$2C$ & $\u\d\d$ & $\textnormal{-}\frac{4}{3\pi}$ & $\frac{4}{\pi}$ & $\frac{4}{3\pi}$ & $1.094(1+\lambda)\tau$ \\
\hline
\end{tabular}
\end{center}
\end{table}

\begin{table}[!t]
\caption{Closed-form expressions of signal on wire 2 and approximate delays ($V_2(L,t) = 1-A_9 e^{-\frac{t}{\tau}} - A_{10}e^{-\frac{t}{(1+2\lambda)\tau}} - A_{11} e^{-\frac{t}{(1+(2-\sqrt{2})\lambda)\tau}} - A_{12} e^{-\frac{t}{(1+(2+\sqrt{2})\lambda)\tau}}$, $\tau_0=\frac{rcL^2}{2}$, and $\tau=\frac{8}{\pi^2}\tau_0$.)}\label{tab:wire2}
\begin{center}
\begin{tabular}{|c|c|p{3mm}|p{3mm}|c|c|c|}
\hline
\multirow{2}{*}{$iC$} & Worst & \multicolumn{4}{|c|}{Coeffs. of $V_2(L,t)$} & \multirow{2}{*}{$T_{b2}^{iC}$}\\
\cline{3-6}
& Pattern & $A_9$ & $A_{10}$ & $A_{11}$ & $A_{12}$ & \\
\hline
$0C$ & $\u\u\u\d$ & $\frac{2}{\pi}$ & $\frac{2}{\pi}$ & $\frac{\sqrt{2}}{\pi}$ & $\textnormal{-}\frac{\sqrt{2}}{\pi}$ & $(\ln \frac{8}{\pi})\tau$ \\
\hline
$1C$ & -$\u\u\d$ & $\frac{1}{\pi}$ & $\frac{3}{\pi}$ & $\frac{\sqrt{2}}{2\pi}$ & $\textnormal{-}\frac{\sqrt{2}}{2\pi}$ & $0.427(1+2\lambda)\tau$\\
\hline
$2C$ & $\d\u\u\d$ & 0 & $\frac{4}{\pi}$ & 0 & 0 & $(\ln \frac{8}{\pi})(1+2\lambda)\tau$ \\
\hline
$3C$ & $\d\u$-$\u$ & $\frac{1}{\pi}$ & $\frac{1}{\pi}$ & $\frac{2-3\sqrt{2}}{2\pi}$ & $\frac{2+3\sqrt{2}}{2\pi}$ & $1.441(1+2\lambda)\tau$\\
\hline
\multirow{2}{*}{$4C$} & \multirow{2}{*}{$\d\u\d\u$} & \multirow{2}{*}{0} & \multirow{2}{*}{0} & \multirow{2}{*}{$\frac{2(1-\sqrt{2})}{\pi}$} & \multirow{2}{*}{$\frac{2(1+\sqrt{2})}{\pi}$} & $6.540(1+$ \\
& & & & & & $(2-\sqrt{2})\lambda)\tau$\\
\hline
\end{tabular}
\end{center}
\end{table}

\subsection{Revised models with consideration of the buffer effects}
\label{sec:revised}
In the previous derivation, the buffer effects are ignored with assumption that the driver resistance and loading capacitance are relatively small. In practice, the values of resistance and capacitance vary with different structure of buffers.
In this work, we consider drivers and receivers implemented as a non-inverting inverter chain. The simplest one has two chained inverters. The loading capacitance $C_L$ and driver resistance $R_S$ are due to the first and last stage inverters in the chain, respectively.
The buffer strength is measured by the normalized size of inverter to the smallest inverter.
For global interconnects in submicron technology, the loading capacitance is not significantly large in comparison with that of interconnect. According to~\cite{ANRS_TCASI08}, for a 45nm technology~\cite{FreePDK45}, the loading capacitance $C_L$ induced by a 100 times inverter is given by 25 fF. In this paper, we consider loading capacitance as large as 100 fF. For significantly large $C_L$, the delay due to $C_L$ would dominate the total propagation delay and all classes of patterns would collapse into one class.
In the following, we revise our models to capture the buffer effects of $R_S$ and $C_L$ at the inputs and outputs of the interconnects, respectively.


First, we focus on our three-wire model. With consideration of buffer effects, the differential equation is still given by Eq.~(\ref{Eq:2}). Only the boundary conditions need to be changed. The revised boundary conditions are given by
\begin{equation*}
\left\{\begin{array}{l}
V_i(0,t) = V_i^p-(V_i^p-V_i^f) u(t)-I_i(0,t) R_S \\
I_i(L,t)=C_L \frac{\partial}{\partial t}V_i(L,t) \quad\quad \mbox{ for } i=1,2,3
\end{array}  \right.
\end{equation*}

By solving the differential equations of a three-wire bus, we derive the expressions of all worst-case patterns as shown in Table~\ref{tab:3dif}.
The revised delay expressions are listed in column five of Table~\ref{tab:3dif}.
Note that the revised three-wire delay model would reduce to that in Table~\ref{tab:3wire}, when the driver resistance and loading capacitance are relatively small, $R_T \doteq 0$ and $C_T \doteq 0$.

\begin{table}[!t]
\caption{Expressions of middle wire in a three-wire model, where $V_2(L,t) = 1 - b_1 B_1 e^{-\frac{t}{\tau_1}} - b_2 B_2 e^{-\frac{t}{\tau_2}}$, $B_1 = 1.01\frac{R_T+C_T+1}{R_T+C_T+\frac{\pi}{4}}$, $B_2 = 1.01\frac{R_T+C_T^\ast+1}{R_T+C_T^\ast+\frac{\pi}{4}}$, $\tau_1 = \frac{RC(R_TC_T+R_T+C_T+(\frac{2}{\pi})^2)}{1.04}$,  $\tau_2 = \frac{(1+3\lambda) RC(R_TC_T^\ast+R_T+C_T^\ast+(\frac{2}{\pi})^2)}{1.04}$,  $R_T=\frac{R_S}{R}$, $C_T=\frac{C_L}{C}$, $C_T^\ast=\frac{C_L}{(1+3\lambda)C}$, $C=cL$, and $R=rL$.} \label{tab:3dif}
\begin{center}
\begin{tabular}{|c|c|c|c|c|}
\hline
\multirow{2}{*}{$iC$} & Worst & \multicolumn{2}{|c|}{Coeffs. of $V_2(L,t)$} & \multirow{2}{*}{$T_3^{iC}$}\\
\cline{3-4}
& Pattern & $b_1$ & $b_2$ & \\
\hline
$0C$ & $\u\u\u$ & 1 & 0 & $(\ln 2B_1)\tau_1$ \\
\hline
$1C$ & $\u\u$- & $\frac{2}{3}$ & $\frac{1}{3}$ & $(\ln 4B_1)\tau_1$ \\
\hline
$2C$ & -$\u$- & $\frac{1}{3}$ & $\frac{2}{3}$ & $(\ln \frac{4B_2}{3})\tau_2$ \\
\hline
$3C$ & $\d\u$- & 0 & 1 & $(\ln 2B_2)\tau_2$ \\
\hline
$4C$ & $\d\u\d$ & $\textnormal{-}\frac{1}{3}$ & $\frac{4}{3}$ & $(\ln \frac{8B_2}{3})\tau_2$ \\
\hline
\end{tabular}
\end{center}
\end{table}

\begin{table}[!t]
\caption{Expressions of middle wire in a five-wire model, where $V_3(L,t) = 1 - b_3 B_3 e^{-\frac{t}{\tau_1}} - b_4 B_4 e^{-\frac{t}{\tau_2}} - b_5 B_5 e^{-\frac{t}{\tau_3}}$, $B_3 = 1.01\frac{R_T+C_T+1}{R_T+C_T+\frac{\pi}{4}}$, $B_4 = 1.01\frac{R_T+C_T^\ast+1}{R_T+C_T^\ast+\frac{\pi}{4}}$, $B_5 = 1.01\frac{R_T+C_T^\dagger+1}{R_T+C_T^\dagger+\frac{\pi}{4}}$, $\tau_1 = \frac{RC(R_TC_T+R_T+C_T+(\frac{2}{\pi})^2)}{1.04}$,  $\tau_2 = \frac{(1+\frac{3}{2}\lambda) RC(R_TC_T^\ast+R_T+C_T^\ast+(\frac{2}{\pi})^2)}{1.04}$, $\tau_3 = \frac{(1+3\lambda) RC(R_TC_T^\dagger+R_T+C_T^\dagger+(\frac{2}{\pi})^2)}{1.04}$,  $R_T=\frac{R_S}{R}$, $C_T=\frac{C_L}{C}$, $C_T^\ast=\frac{C_L}{(1+\frac{3\lambda}{2})C}$, $C_T^\dagger=\frac{C_L}{(1+3\lambda)C}$, $C=cL$, $R=rL$, $f_1 = (\textnormal{-}\ln (\frac{1}{4}+\frac{1}{2} \sqrt{\frac{1}{4}+\frac{3}{2 B_5}}))$, and $f_2 = (\textnormal{-}\ln (\frac{1}{8}+\frac{1}{2}\sqrt{\frac{1}{16}+\frac{3}{2 B_5}}))$.} \label{tab:5dif}
\begin{center}
\begin{tabular}{|c|c|c|c|c|c|}
\hline
\multirow{2}{*}{$iC$} & Worst & \multicolumn{3}{|c|}{Coeffs. of $V_3(L,t)$} & \multirow{2}{*}{$T_5^{iC}$}\\
\cline{3-5}
& Pattern & $b_3$ & $b_4$ & $b_5$ & \\
\hline
$0C$ & $\d\u\u\u\d$ & $\frac{1}{3}$ & $\frac{4}{3}$ & $\textnormal{-}\frac{2}{3}$ & $f_1\tau_3$ \\
\hline
$1C$ & $\d\u\u$-$\d$ & 0 & $\frac{4}{3}$ & $\textnormal{-}\frac{1}{3}$ & $f_2\tau_3$ \\
\hline
$2C$ & $\d$-$\u$-$\d$ & $\textnormal{-}\frac{1}{3}$ & $\frac{4}{3}$ & 0 & $(\ln \frac{8 B_4}{3})\tau_2$ \\
\hline
$3C$ & $\u\d\u$-$\u$ & 0 & 0 & 1 & $(\ln 2 B_5)\tau_3$ \\
\hline
$4C$ & $\u\d\u\d\u$ & $\textnormal{-}\frac{1}{3}$ & 0 & $\frac{4}{3}$ & $(\ln \frac{8 B_5}{3})\tau_3$ \\
\hline
\end{tabular}
\end{center}
\end{table}


Similarly, for a five-wire bus, we derive the expressions of all worst-case patterns in each class as shown in Table~\ref{tab:5dif}.
The ratio between $\tau_2$ and $\tau_3$ is given by $\frac{\tau_2}{\tau_3}=\frac{(1+\frac{3}{2}\lambda) RC (R_TC_T^\ast+R_T+C_T^\ast+(\frac{2}{\pi})^2) }{(1+3\lambda) RC(R_TC_T^\dagger+R_T+C_T^\dagger+(\frac{2}{\pi})^2)} \doteq \frac{1}{2}$. Then the 50\% delay can be solved by assuming $e^{-\frac{t}{\tau_2}} = (e^{-\frac{t}{\tau_3}})^2$. The revised delay expressions are listed in column six of Table~\ref{tab:5dif}.
Note that the revised five-wire delay model would reduce to that in Table~\ref{tab:5wire}, when the driver resistance and loading capacitance are relatively small, $R_T \doteq 0$ and $C_T \doteq 0$.

\begin{table}[!t]
\caption{Bus parameters in a 45$\mathrm{nm}$ technology.} \label{tab:parameter}
\begin{center}
\begin{tabular}{|c|c||c|c|}
\hline
\multicolumn{4}{|c|}{Parameters} \\
\hline
$L$ & 5 mm & $r$ & 13.75 $\Omega$/mm \\
\hline
$w$ & 0.8 $\mu$m & $l$ & 1.736 nH/mm \\
\hline
$s$ & 0.8 $\mu$m & $c$ & 8.263 fF/mm \\
\hline
$t$ & 2 $\mu$m & $c_c$ & 101.136 fF/mm \\
\hline
$h$ & 4.82 $\mu$m & $R_S$ & 100 $\Omega$ \\
\hline
$K_{\textrm{ILD}}$ & 2.5 & $C_L$ & 0 fF \\
\hline
\end{tabular}
\end{center}
\end{table}

According to the delay expressions in Tables~\ref{tab:3dif} and \ref{tab:5dif}, both driver resistance and loading capacitance tend to increase the delay. When the loading capacitance increases, the delay difference among all classes diminishes. For extremely large $C_L$, the delays for all classes are close and the classification becomes inconsequential.

\subsection{Characterization of the delay of a multi-wire bus}
\label{sec:application}

In the derivation of our five-wire model above, we focus on the worst-case patterns of the middle wires only.
We also derive delay models for boundary wires. In the following, we show that our five-wire model can be easily applied to approximate the delays of an $m$-wire bus ($m>5$).
First, we use our five-wire delay model as a shift window to scan the internal wires (wires 3 through $m-2$) to identify the longest delay.
Then, for boundary wires (wires 1, 2, $m-1$, and $m$), we use the models in Tables~\ref{tab:wire1} and \ref{tab:wire2} for delay approximation. Hence, the delay of an $m$-wire bus is given by the largest delay among all wires.
For example, for a pattern $\u\d\u\d\d\d$ of a six-wire bus, the classes for wires 1 through 6 are given by $2C$, $4C$, $4C$, $2C$, $0C$, and $0C$, respectively. Thus, the worst-case class is given by $4C$. According to our models in Tables~\ref{tab:5wire}, \ref{tab:wire1}, and \ref{tab:wire2}, the worst-case delay is given by the larger one of the two delays $6.540(1+(2-\sqrt{2})\lambda)\tau$ and $\left(\ln{\frac{32}{3\pi}}\right)(1+3\lambda)\tau$.

The proposed analytical delay models target two important applications. One primary application of our model is the design of crosstalk avoidance codes (CACs). Since our proposed models provide more accurate delays for different transition patterns than previous models, we can identify unwanted patterns more effectively. Second, our models can be applied to partial coding schemes, where buses are broken into sub-buses, since our models are more accurate for a bus of small size.
To incorporate such analytical delay models in EDA softwares, such as a typical timing analysis flow, appropriate adjustments are needed. We plan to investigate this important scenario in our future work.

\subsection{Discussion on synchronization problems}
\label{sec:sync}

In previous subsections, we assume simultaneous transitions on all the wires. However, for global buses where buffer insertion techniques are usually used to reduce their delay \cite{Tsa06}, simultaneous signal transitions on the bus cannot be guaranteed. Our derived models do not work for buses with synchronization problems.
In the following, we briefly discuss the synchronization problems and conclude with insights on the delay changes of interconnects with synchronization problems and impacts on the CAC designs.

Based on our three-wire and five-wire models, we observe \textbf{two possible scenarios} with regard to the impact of synchronization problems on the delay. When the time differences are relatively small, the delay is increased only by the time differences. When the time differences are sufficiently large, they can change the worst delay of a class to a different class. For instance, the delays of the transition patterns in $0C$ and $1C$ may be increased to those of $2C$ when the time differences are large enough, and similarly $2C$ to $3C$. This is consistent with the observation in~\cite{Dua04}. On the other hand, the delays of the transition patterns in $3C$ may decrease to those of $2C$ class when the time differences are large enough. Intuitively, this is because large time differences change the intended transition patterns into different patterns.
As observed above, depending on the severity of the synchronization problems, the effectiveness of CACs is affected to a varying extent. Furthermore, the sensitiveness to time differences varies with CACs.


\section{Performance evaluation}\label{sec:simulation}
We evaluate the performance of our delay models, and compare it with that of the model in \cite{Sot01} in three scenarios. First, since our delay models focus on three and five adjacent wires, we consider three- and five-wire buses. This scenario is also motivated by partial coding schemes (see, e.g., \cite{Dua01}, \cite{Vic01}, and \cite{Sri07}), which divide a wide bus into sub-buses with a few wires and separate them by shielding wires.
The second scenario is buses with more than five wires. We have run extensive simulations on buses with an odd number of wires (up to 33 wires). Our conclusions are the same regardless of the number of wires. For brevity, we present our simulation results for 17- and 33-wire buses.
In the first two scenarios, we focus on the worst-case delays of the middle wires. To characterize the whole bus transitions, our five-wire model can be applied to all wires to approximate their delays with higher accuracy.
In the third scenario, we assume the transition patterns are limited to those of CACs and consider the worst-case delays for all wires of an 8-wire bus.

\begin{figure}[!tb]
\begin{minipage}[b]{1.0\linewidth}
  \centering
 \centerline{\epsfig{figure=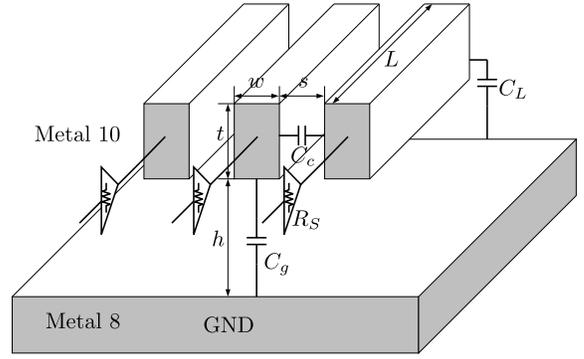,width=7.5cm}}
\end{minipage}
\caption{Interconnect structure.}
\label{fig:interconnect}
\end{figure}

All the simulation results in this paper are obtained by the following setup. The simulation is based on a 45nm technology with 10 metal layers~\cite{FreePDK45}. The global buses are routed in the top two metal layers, 10 and 9, with a ground metal layer 8 down below as shown in Fig.~\ref{fig:interconnect}.
We consider metal layer 10 for all buses, since the crosstalk is more serious than that of metal layer 9.
The bus parameters are obtained by structure 1 in \cite{PTM} and summarized in Table~\ref{tab:parameter}, where $K_{\textrm{ILD}}$ is the permittivity of the dielectric between metals.
Since the model in~\cite{Sot01} does not account for the loading capacitance, we assume $C_L = 0$ fF for simulations in comparison with the model in~\cite{Sot01}. We also simulate 17- and 33-wire buses with $C_L = 100$ fF, which represents the loading capacitance induced by a 400 times inverter.
The coupling factor is given by $\lambda = \frac{c_c}{c} \doteq 12.2$.
For inputs with $t_r = 10$ ps, inductance effect is negligible when 1.3 mm $< L <$ 66.7 mm. All the buses for simulation have a length of 5 mm and the inductance effect is not considered in this work. The buses are divided into 100 sections as shown in Fig.~\ref{fig:3bus} to characterize the distributed RC model. The simulation results are obtained from HSPICE.


\subsection{Three-wire and five-wire buses}
For a three-wire bus, the simulated delays are compared with the delays by our model and the model in \cite{Sot01} for all classes in Table~\ref{tab:1}, where $T_d$ denotes the simulated worst delay of wire 2, $T^{iC}_3$ the approximate delay for $iC$ pattern by our three-wire model, and $T_2$ by the model in \cite{Sot01}. The error percentages of our model and the model in \cite{Sot01} are also shown in Table~\ref{tab:1}. For all five classes of transition patterns, the \textbf{maximum} and \textbf{minimum} errors by our model are only 3.14\% and 0.47\%, respectively, as opposed to 891.90\% and 34.38\% by the model in \cite{Sot01}, respectively. As Table~\ref{tab:1} shows, our model is much more accurate than the model in \cite{Sot01} for all patterns in a three-wire bus. We remark that the delay by our model for the $1C$ pattern, $\left(\ln{\frac{16}{\pi}}\right)\tau$, does not depend on $\lambda$.

\begin{table}
\caption{Comparison of simulated delays, delays of our three-wire model and the model in \cite{Sot01}. All the delays are in $\mathrm{ps}$.}\label{tab:1}
\begin{center}
\begin{tabular}{|r|c|c|c|c|c|c|}
\hline
\multirow{2}{*}{$iC$}& Worst & Sim. & \multicolumn{2}{|c|}{Our model} & \multicolumn{2}{|c|}{\cite{Sot01}}\\
\cline{3-7} & pattern & $T_d$ & $T^{iC}_3$ & $\frac{|T^{iC}_3-T_d|}{T_d}$ & $T_2$ &  $\frac{|T_2-T_d|}{T_d}$\\
\hline
$0C$ & $\uparrow \uparrow \uparrow$ & 3.96 & 4.04 & 2.02\% & 5.55 & 40.15\%\\
\hline
$1C$ & $\uparrow \uparrow$- & 7.41 & 7.56 & 2.02\% & 73.50  & 891.90\%\\
\hline
$2C$ & -$\uparrow$- & 72.28 & 74.55 &  3.14\% & 141.45 & 95.70\%\\
\hline
$3C$ & $\downarrow \uparrow$- & 150.74 & 152.24 &  1.00\% & 209.40 & 38.91\%\\
\hline
$4C$ & $\downarrow \uparrow \downarrow$ & 206.40 & 207.36 & 0.47\% & 277.35  & 34.38\%\\
\hline
\end{tabular}
\end{center}
\end{table}

\begin{table}
\caption{Comparison of simulated delays, delays of our five-wire model and the model in \cite{Sot01}. All the delays are in $\mathrm{ps}$.}\label{tab:2}
\begin{center}
\begin{tabular}{|r|c|c|c|c|c|c|}
\hline
\multirow{2}{*}{$iC$}& Worst & Sim. & \multicolumn{2}{|c|}{Our model} & \multicolumn{2}{|c|}{\cite{Sot01}}\\
\cline{3-7} & pattern & $T_d$ & $T^{iC}_5$ & $\frac{|T^{iC}_5-T_d|}{T_d}$ & $T_3$ &  $\frac{|T_3-T_d|}{T_d}$\\
\hline
$0C$ & $\downarrow \uparrow \uparrow \uparrow \downarrow$ & 35.30 & 23.15 & 34.41\% & 5.55 & 84.28\%\\
\hline
$1C$ & $\downarrow$-$\uparrow \uparrow \downarrow$ & 63.09 & 62.09 & 1.59\% & 73.50  & 16.50\%\\
\hline
$2C$ & $\downarrow$-$\uparrow$-$\downarrow$ & 98.39 & 106.43 &  8.17\% & 141.45 & 43.76\%\\
\hline
$3C$ & $\uparrow$-$\uparrow \downarrow \uparrow$ & 134.19 & 152.24 &  13.45\% & 209.40 & 56.05\%\\
\hline
$4C$ & $\uparrow \downarrow \uparrow \downarrow \uparrow$ & 218.91 & 207.36 & 5.28\% & 277.35  & 26.70\%\\
\hline
\end{tabular}
\end{center}
\end{table}

\begin{table*}
\caption{Comparison of simulated delays and delays given by our five-wire model and the model in \cite{Sot01} for wire 9 in a 17-wire bus with $C_L = 0$ $\mathrm{fF}$. All the delays are in $\mathrm{ps}$.}\label{tab:5}
\begin{center}
\begin{tabular}{|c|c|c|c|c|c|c|}
\hline
\multirow{3}{*}{$iC$} & Worst patterns & Sim. & \multicolumn{2}{c|}{Our model} & \multicolumn{2}{c|}{\cite{Sot01}} \\
\cline{3-7}
& via Alg.~\ref{alg:findworst} & $T_d$ & $T^{iC}_5$ & $\frac{|T^{iC}_5-T_d|}{T_d}$ & $T_9$ &  $\frac{|T_9-T_d|}{T_d}$ \\
\hline
$0C$ & $\uparrow \uparrow \uparrow \uparrow \downarrow \downarrow \downarrow (\uparrow \uparrow \uparrow) \downarrow \downarrow \downarrow \uparrow \uparrow \uparrow \uparrow$ & 42.17 &  23.15 & 45.10\% &  5.55  & 86.84\% \\
\hline
$1C$ & $\uparrow \uparrow \uparrow \uparrow \uparrow \downarrow \downarrow (\uparrow \uparrow \mbox{-}) \downarrow \downarrow \uparrow \uparrow \uparrow \uparrow \uparrow$ & 67.50 & 62.09 & 8.01\% & 73.50  & 8.89\% \\
\hline
$2C$ & $\downarrow \downarrow \uparrow \uparrow \uparrow \uparrow \downarrow (\mbox{-} \uparrow \mbox{-}) \downarrow \uparrow \uparrow \uparrow \uparrow \downarrow \downarrow$ & 112.82 & 106.43 &  5.66\% & 141.45 & 25.38\% \\
\hline
$3C$ & $\downarrow \downarrow \downarrow \uparrow \uparrow \uparrow \downarrow (\downarrow \uparrow \mbox{-}) \downarrow \uparrow \uparrow \uparrow \downarrow \downarrow \downarrow$ & 165.44 & 152.24 &  7.98\% & 209.40 & 26.57\% \\
\hline
$4C$ & $\uparrow \downarrow \downarrow \downarrow \uparrow \uparrow \uparrow (\downarrow \uparrow \downarrow) \uparrow \uparrow \uparrow \downarrow \downarrow \downarrow \uparrow$ & 228.46 & 207.36 & 9.24\% & 277.35  & 21.40\% \\
\hline
\end{tabular}
\end{center}
\end{table*}

\begin{table*}
\caption{Comparison of simulated delays and delays given by our five-wire model and \cite{Sot01} for wire 17 in a 33-wire bus with $C_L = 0$ $\mathrm{fF}$. All the delays are in $\mathrm{ps}$.}\label{tab:6}
\begin{center}
\begin{tabular}{|c|c|c|c|c|c|c|}
\hline
\multirow{2}{*}{$iC$} & Worst patterns & Sim. & \multicolumn{2}{c|}{Our model} & \multicolumn{2}{c|}{\cite{Sot01}} \\
\cline{3-7}
& via Alg.~\ref{alg:findworst} & $T_d$ & $T^{iC}_5$ & $\frac{|T^{iC}_5-T_d|}{T_d}$ & $T_{17}$ &  $\frac{|T_{17}-T_d|}{T_d}$ \\
\hline
$0C$ & $\downarrow \downarrow \downarrow \downarrow \uparrow \uparrow \uparrow \uparrow \uparrow \uparrow \uparrow \uparrow \downarrow \downarrow \downarrow (\uparrow \uparrow \uparrow) \downarrow \downarrow \downarrow \uparrow \uparrow \uparrow \uparrow \uparrow \uparrow \uparrow \uparrow \downarrow \downarrow  \downarrow \downarrow$ & 42.27 &  23.15 & 45.23\% & 5.55  & 86.87\% \\
\hline
$1C$ & $\downarrow \downarrow \downarrow \downarrow \downarrow \downarrow \downarrow \uparrow \uparrow \uparrow \uparrow \uparrow \uparrow \downarrow \downarrow (\mbox{-}\uparrow\uparrow) \downarrow \downarrow \uparrow \uparrow \uparrow \uparrow \uparrow \uparrow \downarrow \downarrow \downarrow \downarrow \downarrow \downarrow \downarrow$ & 68.30 & 62.09 & 9.09\% & 73.50  & 7.61\% \\
\hline
$2C$ & $\uparrow \uparrow \uparrow \uparrow \downarrow \downarrow \downarrow \downarrow \downarrow \downarrow \uparrow \uparrow \uparrow \uparrow \downarrow (\mbox{-} \uparrow \mbox{-}) \downarrow \uparrow \uparrow \uparrow \uparrow \downarrow \downarrow \downarrow \downarrow \downarrow \downarrow \uparrow \uparrow \uparrow \uparrow$ & 113.16 & 106.43 &  5.95\% & 141.45 & 25.00\% \\
\hline
$3C$ & $\downarrow \downarrow \uparrow \uparrow \uparrow \uparrow  \uparrow \downarrow \downarrow \downarrow \downarrow \uparrow \uparrow \uparrow \downarrow (\mbox{-} \uparrow \downarrow) \downarrow \uparrow \uparrow \uparrow \downarrow \downarrow \downarrow \downarrow \uparrow  \uparrow \uparrow \uparrow \uparrow \downarrow \downarrow$ & 165.57 & 152.24 &  8.05\% & 209.40 & 26.47\% \\
\hline
$4C$ & $\downarrow \downarrow \downarrow\downarrow \uparrow \uparrow \uparrow \uparrow \downarrow \downarrow \downarrow \downarrow \uparrow \uparrow \uparrow (\downarrow \uparrow \downarrow) \uparrow \uparrow \uparrow \downarrow \downarrow \downarrow \downarrow \uparrow \uparrow \uparrow \uparrow \downarrow \downarrow \downarrow \downarrow$  & 229.02 & 207.36 & 9.46\% & 277.35 & 21.10\% \\
\hline
\end{tabular}
\end{center}
\end{table*}

\begin{table*}
\caption{Comparison of simulated delays and delays given by our five-wire model focusing on the middle wire in a 17-wire and a 33-wire buses with $C_L = 100$ $\mathrm{fF}$. All the delays are in $\mathrm{ps}$.}\label{tab:100fF}
\begin{center}
\begin{tabular}{|c|c|c|c|c|c|c|c|c|}
\hline
\multirow{2}{*}{$iC$}& Worst 17-wire patterns & Sim. & \multicolumn{2}{c|}{Our model} & Worst 33-wire patterns & Sim. & \multicolumn{2}{c|}{Our model} \\
\cline{3-5}\cline{7-9}
& via Alg.~\ref{alg:findworst} & $T_d$ & $T^{iC}_5$ & $\frac{|T^{iC}_5-T_d|}{T_d}$ & via Alg.~\ref{alg:findworst} & $T_d$ & $T^{iC}_5$ & $\frac{|T^{iC}_5-T_d|}{T_d}$\\
\hline
$0C$ & $\uparrow \uparrow \uparrow \downarrow \downarrow \downarrow \downarrow (\uparrow \uparrow \uparrow) \downarrow \downarrow \downarrow \downarrow \uparrow \uparrow \uparrow$ & 50.75 &  25.11 & 50.52\% & $\downarrow \uparrow \uparrow \uparrow \uparrow \uparrow \uparrow \uparrow \uparrow \uparrow \uparrow \downarrow \downarrow \downarrow \downarrow (\uparrow \uparrow \uparrow) \downarrow \downarrow \downarrow \downarrow \uparrow \uparrow \uparrow \uparrow \uparrow \uparrow \uparrow \uparrow \uparrow  \uparrow \downarrow$ & 50.78 &  25.11 & 50.55\% \\
\hline
$1C$ & $\uparrow \uparrow \uparrow \uparrow \downarrow \downarrow \downarrow (\uparrow \uparrow \mbox{-}) \downarrow \downarrow \downarrow \uparrow \uparrow \uparrow \uparrow$ & 76.42 & 67.35 & 11.87\% & $\downarrow \downarrow \downarrow \downarrow \downarrow \uparrow \uparrow \uparrow \uparrow \uparrow \uparrow \uparrow \downarrow \downarrow \downarrow (\mbox{-}\uparrow\uparrow) \downarrow \downarrow \downarrow \uparrow \uparrow \uparrow \uparrow \uparrow \uparrow \uparrow \uparrow \downarrow \downarrow \downarrow \downarrow$ & 76.43 & 67.35 & 11.88\% \\
\hline
$2C$ & $\uparrow \uparrow \uparrow \uparrow \uparrow \uparrow \downarrow (\mbox{-} \uparrow \mbox{-}) \downarrow \uparrow \uparrow \uparrow \uparrow \uparrow \uparrow$ & 118.92 & 123.46 &  3.82\% & $\uparrow \downarrow \downarrow \downarrow \downarrow \downarrow \downarrow \downarrow \uparrow \uparrow \uparrow \uparrow \uparrow \uparrow \downarrow (\mbox{-} \uparrow \mbox{-}) \downarrow \uparrow \uparrow \uparrow \uparrow \uparrow \uparrow \downarrow \downarrow \downarrow \downarrow \downarrow \downarrow \downarrow \uparrow$ & 119.21 & 123.46 &  3.57\% \\
\hline
$3C$ & $\downarrow \downarrow \uparrow \uparrow \uparrow \uparrow \downarrow (\downarrow \uparrow \mbox{-}) \downarrow \uparrow \uparrow \uparrow \uparrow \downarrow \downarrow$ & 177.71 & 164.62 &  7.37\% & $\downarrow \downarrow \uparrow \uparrow \downarrow \downarrow  \downarrow \downarrow \downarrow \downarrow \uparrow \uparrow \uparrow \uparrow \downarrow (\mbox{-} \uparrow \downarrow) \downarrow \uparrow \uparrow \uparrow \uparrow \downarrow \downarrow \downarrow \downarrow  \downarrow \downarrow \uparrow \uparrow \downarrow \downarrow$ & 177.74 & 164.62 &  7.38\% \\
\hline
$4C$ & $\uparrow \downarrow \downarrow \uparrow \uparrow \uparrow \uparrow (\downarrow \uparrow \downarrow) \uparrow \uparrow \uparrow \uparrow \downarrow \downarrow \uparrow$ & 236.18 & 224.41 & 4.98\% & $\downarrow \downarrow \uparrow \uparrow \uparrow \uparrow \uparrow \downarrow \downarrow \downarrow \downarrow \uparrow \uparrow \uparrow \uparrow (\downarrow \uparrow \downarrow) \uparrow \uparrow \uparrow \uparrow \downarrow \downarrow \downarrow \downarrow \uparrow \uparrow \uparrow \uparrow \uparrow \downarrow \downarrow$  & 236.67 & 224.41 & 5.18\% \\
\hline
\end{tabular}
\end{center}
\end{table*}

For a five-wire bus, the worst delays of all classes of transition patterns based on our five-wire model are compared with those of the model in \cite{Sot01} as well as the simulated delays by HSPICE in Table~\ref{tab:2}, where $T_d$ denotes the simulated worst-case delay of wire 3 for all $iC$ patterns, $T^{iC}_5$ the approximate delay for $iC$ pattern by our five-wire model, and $T_3$ by the model in \cite{Sot01}. The error percentages of our model and the model in \cite{Sot01} are shown in Table~\ref{tab:2}. For a five-wire bus the \textbf{maximum} and \textbf{minimum} errors by our model are 34.41\% and 1.59\%, respectively, in comparison to 84.28\% and 16.50\% by the model in \cite{Sot01}, respectively.
As Table~\ref{tab:2} shows, our five-wire model is more accurate than the model in \cite{Sot01} for all patterns in a five-wire bus. In particular, although the delays in the model in \cite{Sot01} were claimed to be upper bounds on the actual delays, our simulation results in Table~\ref{tab:2} show that this claim is invalid for the $0C$ patterns. In \cite{Dua04}, the author proposed a method which achieves a delay of $\tau_0$ by surrounding each data wire with two shield wires with the same transition. Since the transition patterns for each data wire are always in $0C$ class, the delays of the data wires are $\tau_0$ according to the model in \cite{Sot01}. In contrast, the delay for the data wires can be as large as $0.165(1+3\lambda)\tau_0$ by our model; When $\lambda$ is large, the model in \cite{Sot01} severely underestimates the delay, while our model is more accurate.


\subsection{17-wire and 33-wire buses}
We next compare our five-wire model with the model in \cite{Sot01} for 17- and 33-wire buses. With a 17-wire bus, we focus on the middle wire (wire 9). We still classify the transition patterns according to the transitions of the middle three wires (wires 8, 9, and 10).
Since it is time consuming to identify the transition patterns with the longest delay in each class, we make one assumption about the patterns with the longest delay in each class. For any two wires symmetric to wire 9 (wire $i$ and wire $18$-$i$, $i \in \{1,2,\cdots, 8\}$), there are nine possible patterns, $\uparrow \uparrow$, $\downarrow \downarrow$, - -, $\uparrow$-, -$\uparrow$, $\downarrow$-, -$\downarrow$, $\uparrow \downarrow$, and $\downarrow \uparrow$. For patterns in opposite direction, we assume the influences of the two wires will cancel out because of symmetry. For other patterns, if the upward transition of one wire increases the delay, we see that $\uparrow \uparrow$ has greater delay than $\uparrow$- or -$\uparrow$. Similarly, if the downward transition increases the delay, the pattern $\downarrow \downarrow$ has greater delay than $\downarrow$- or -$\downarrow$. So we assume that the longest delay happens when two symmetric wires have either $\uparrow \uparrow$ or $\downarrow \downarrow$ transitions.

Based on this assumption, we search all possible symmetric transition patterns to find the worst-case patterns in each class, which
are listed in the second column of Table~\ref{tab:5}, where the pattern on wires 8, 9, and 10 are shown in the parenthesis. The simulated worst-case delays for all $iC$, denoted by $T_d$, are compared with the delays by our five-wire model and the model in \cite{Sot01} in Table~\ref{tab:5}. The error percentage of our model and the model in \cite{Sot01}  are also shown in Table~\ref{tab:5}. For all five classes, the \textbf{maximum} and \textbf{minimum} errors by our model are only 45.10\% and 5.66\%, respectively, as opposed to 86.84\% and 8.89\% by the model in \cite{Sot01}, respectively. For all classes except $1C$, our five-wire model outperforms the model in \cite{Sot01}. The model in \cite{Sot01} also has a large error percentage for $0C$.

With a 33-wire bus, we focus on the delay of the middle wire (wire 17). Since there are $3^{33}$ transition patterns, it is infeasible to search all possible symmetric transitions as before to find the worst-case patterns. We make the following three assumptions: (1) The worst patterns in each classes are symmetric; (2) The closer the wire gets to the middle wire, the greater the coupling on the settling of the middle wire; (3) We initialize the middle three wires to a pattern in $iC$, and initialize all other wires with opposite transitions to the middle wire. Based on these three assumptions, we use Alg.~\ref{alg:findworst} to find the patterns with largest delays. We denote by $P_i$ the updated transition pattern of an $m$-wire bus after the $i$-th iteration of Alg.~\ref{alg:findworst}, where $m$ is odd. Alg.~\ref{alg:findworst} can greatly reduce the simulation time for identifying the worst-case patterns. For instance, the worst-case patterns for an $33$-wire bus can be identified by simulating only $5 \times 15 = 75$ transition patterns.


\begin{algorithm}[!tp]
  \caption{The algorithm for identifying the worst-case pattern, with respect to the three assumptions, in an $m$-wire bus.}
  \begin{algorithmic}
    \REQUIRE $m$-wire bus;
    \STATE \textbf{Initialize}: $P_0$ is initialized with transitions opposite to wire $\frac{m+1}{2}$, except for wires $\frac{m-1}{2}$, $\frac{m+1}{2}$, and $\frac{m+3}{2}$;
    \STATE $i=0$;
    \REPEAT
    \FOR{$j$ = $\frac{m-3}{2}$ to 1}
    \STATE Flip the transition of wires $j$ and ($m+1-j$) in $P_i$;
    \IF{the delay of wire $\frac{m+1}{2}$ increases}
    \STATE Keep the changes;
    \ELSE
    \STATE Reverse the changes;
    \ENDIF
    \ENDFOR
    \STATE $i=i+1$;
    \STATE Update $P_{i}$ with the current pattern;
    \UNTIL{$P_{i-1} = P_i$}
    \RETURN Worst-case transition pattern for wire $\frac{m+1}{2}$;
  \end{algorithmic}
  \label{alg:findworst}
\end{algorithm}

We note that the one assumption about the worst-case patterns for 17-wire buses and the three assumptions about 33-wire buses are made in order to reduce the complexity of finding the worst-case patterns. We did verify our three assumptions about 33-wire buses over 9- and 11-wire buses: the worst cases for all the classes based on Alg.~\ref{alg:findworst} are indeed the worst cases by exhaustive search.
This also verifies the assumption for 17-wire buses, since it is one of the three assumptions for 33-wire buses.
For instance, the worst-case $2C$ pattern of a 11-wire bus is given by $\u\u\u\d$-$\u$-$\d\u\u\u$ with exhaustive search. In Alg.~\ref{alg:findworst}, starting from $\d\d\d\d$-$\u$-$\d\d\d\d$, the worst-case pattern is found via the order: $\d\d\d\d$-$\u$-$\d\d\d\d$ $\Longrightarrow$ $\d\d\u\d$-$\u$-$\d\u\d\d$ $\Longrightarrow$ $\d\u\u\d$-$\u$-$\d\u\u\d$ $\Longrightarrow$ $\u\u\u\d$-$\u$-$\d\u\u\u$.
Unfortunately, it is difficult to verify Alg.~\ref{alg:findworst}, even for one case, for 17- or 33-wire buses, because the complexity would be prohibitive. For instance, for each class focusing on the middle wire, there are $3^{14}=4782969$ possible patterns for a 17-wire bus (and $3^{30}\doteq 2.06\times 10^{14}$ for a 33-wire bus), and it takes about 166 days to simulate these cases.

The worst transition patterns for each class in a 33-wire bus, with respect to the three assumptions above, are listed in the second column of Table~\ref{tab:6}, where the pattern on wires 16, 17, and 18 are shown in the parenthesis.
The simulated worst-case delays of wire 17 for all $iC$ patterns, denoted by $T_d$, are compared with the delays of our five-wire model and the model in \cite{Sot01}. The error percentages of our model and the model in \cite{Sot01} are also shown in Table~\ref{tab:6}. The \textbf{maximum} and \textbf{minimum} errors by our model are only 45.23\% and 5.95\%, respectively, in comparison to 86.87\% and 7.61\% by the model in \cite{Sot01}, respectively. Again, for all classes except $1C$, our five-wire model outperforms the model in \cite{Sot01}. The model in \cite{Sot01} also has a large error percentage for $0C$.

Since our revised models also account for the loading capacitance, we also simulate 17- and 33-wire buses with $C_L = 100$ fF, which represents the loading capacitance induced by a 400 times inverter. The simulated worst-case delays of the middle wire for all $iC$ patterns, denoted by $T_d$, are compared with the delays $T_5^{iC}$ by our five-wire model as shown in Table~\ref{tab:100fF}. The error percentages of our model are also shown in Table~\ref{tab:100fF}. The worst-case patterns are obtained via Alg.~\ref{alg:findworst}. The worst-case patterns are different from those in Tables \ref{tab:5} and \ref{tab:6} due to the varying of the loading capacitances. However, our five-wire model can still approximate the delays with similar error percentages as those in Tables \ref{tab:5} and \ref{tab:6}.

Finally, we remark that the longest delays for each class in Tables \ref{tab:5} and \ref{tab:6} are approximately the same for both 17- and 33-wire buses.
Based on the simulation results of 17- and 33-wire buses, we conjecture that our five-wire model would be more accurate than the model in \cite{Sot01} for buses with any number of wires.

\subsection{Performance of CACs}\label{sec:simCACs}
In the simulation results above, we assume the transition patterns are arbitrary. Herein, we assume the transition patterns are limited to those of CACs. We evaluate the performance of our delay model for three families of CACs \cite{Dua01,Vic01,Sri07}: {\it one Lambda codes} (OLCs), {\it forbidden pattern codes} (FPCs), and {\it forbidden overlap codes} (FOCs).
Based on our five-wire model, the worst delays of aforementioned CACs are shown in Tables~\ref{tab:5wire}, \ref{tab:wire1}, and \ref{tab:wire2}.
Based on the model in \cite{Sot01}, the worst delays of aforementioned CACs are approximated by $(1+\lambda)\tau_0$, $(1+2\lambda)\tau_0$, and $(1+3\lambda)\tau_0$, respectively.
Since the number of transition patterns is a quadratic function of the number of codewords, it is time-consuming to simulate a large bus to get the worst-case delays on all wires.
Hence, for each CAC, we simulate an 8-wire bus. The numbers of codewords of OLC, FPC, and FOC are given by 16, 68, and 149, respectively. The total numbers of transition patterns for OLC, FPC, and FOC are given by 240, 4556, and 22052, respectively.
We obtain by simulation the maximum delays of each wire for all transition patterns. The simulation results are shown in Table~\ref{tab:8wire}, where the delays given by our five-wire model and the model in \cite{Sot01} are also included. Intuitively, the worst-case delays of any two symmetric wires are the same, since the symmetric transition of a valid transition pattern is also valid.
As shown in Table~\ref{tab:8wire}, the simulated delays of symmetric wires are very close.
For OLCs, FPCs, and FOCs, the largest delays are emphasized in boldface. As Table~\ref{tab:8wire} shows,
our delay models are more accurate than the model in \cite{Sot01} for all three families of CACs.

\begin{table}[!t]
\caption{Comparison of simulated delays and delays given by our five-wire model and \cite{Sot01} for all wire in an 8-wire bus, where $T^{iC}_5$, $T^{iC}_{b1}$, and $T^{iC}_{b2}$ denote the delays of wires 3-8, wire 1 ($m$), and wire 2 ($m-1$), respectively. All the delays are in $\mathrm{ps}$.}\label{tab:8wire}
\begin{center}
\begin{tabular}{|c|c|c|c|}
\hline
\multirow{2}{*}{wire $i$} & \multicolumn{3}{c|}{Delays}\\
\cline{2-4}
& OLC & FPC& FOC\\
\hline
1 & \textbf{55.36} & \textbf{107.43} & 107.73\\
\hline
2 & 32.20 & 102.71 & 159.65\\
\hline
3 & 51.40 & 106.65 & 154.59\\
\hline
4 & 51.06 & 101.91 & 162.61\\
\hline
5 & 50.79 & 101.89 & \textbf{162.77}\\
\hline
6 & 51.39 & 106.53 & 154.62\\
\hline
7 & 32.46 & 102.72 & 160.61\\
\hline
8 & \textbf{55.36} & 107.39 & 108.88\\
\hline
\hline
\cite{Sot02} & 73.50 & 141.45 & 209.40\\
\hline
$T^{iC}_5$ & 62.09 & 106.43 & 152.24\\
\hline
$T^{iC}_{b1}$ & 53.43 & 98.76 & 98.76 \\
\hline
$T^{iC}_{b2}$ & 42.52 & 102.84 & 157.64\\
\hline
\end{tabular}
\end{center}
\end{table}

\section{Conclusions and future work}\label{sec:conclusion}
In this paper, we propose improved analytical delay models for coupled interconnects. We first derive closed-form expressions of the signals on the bus, based on the distributed RC model, and then approximate the delays of different patterns by evaluating these closed-form expressions. We focus on three-wire and five-wire models, and simulation results show that our model has better accuracy than the model in \cite{Sot01}. Although our models are based on three-wire and five-wire buses, they are not limited to these two cases. For a bus with more than five wires, our five-wire model can still approximate delays better than the model in \cite{Sot01}. 
\bibliographystyle{IEEEtran}
\bibliography{bibbus}

\end{document}